\documentclass[prb,aps,showpacs,twocolumn]{revtex4-1}
\usepackage{bm}
\usepackage{graphicx}
\usepackage{epsfig}
\begin{document}

\title{Effect of carrier heating on  photovoltage in FET}
\author{E.L.~Ivchenko}\email{ivchenko@coherent.ioffe.ru}
\affiliation{Ioffe Physico-Technical Institute, 194021 St.~Petersburg, Russia }
\begin{abstract}
We have calculated, within the framework of the Boltzmann equation, a dc electric current and emf in a two-dimensional system induced by the high-frequency field of an electromagnetic wave or the electric field of a plasmon wave. It is established that the generated current consists of two contributions, one of which is proportional to the real part of the wave vector projection of the exciting wave onto the interface plane and represents the electron drag effect, and the other contribution is proportional to the extinction coefficient of the wave in the interface plane. It is shown that the main cause of the second contribution is a nonuniform electron heating created by the wave and controlled by the energy relaxation time of the electron gas. In FET the heating mechanism of the electric-current formation can remarkably exceed the current calculated neglecting the heating.
\end{abstract}


\maketitle

\section{Introduction}
Field-effect transistors (FETs) can be used as effective detectors of terahertz radiation, as has been shown theoretically by Dyakonov and Shur [\onlinecite{Shur},\onlinecite{Shur2}] and demonstrated
in numerous experimental studies, see e.g. Refs.~[\onlinecite{knap, knap2009, ganichev2012}]. The operation of such detectors is based on the generation of emf in the transistor two-dimensional channel as a result of propagation along the channel of the plasmon wave excited by terahertz radiation. In this paper we show that, in addition to the contribution to the photo-emf, calculated previously [\onlinecite{Shur2}], there exists a contribution which exceeds the previous one, arises due to the nonuniform heating of carriers by the plasmon wave and depends on the energy relaxation time of the electron gas.
\section{The dc current for the constant collision rate}
Here we obtain an expression for the dc current in a two-dimensional system induced by a high-frequency electromagnetic field. Given that the detailed derivation for a three-dimensional system is presented in Ref.~[\onlinecite{PerPinsk}], we omit most of the intermediate steps. Following Perel and Pinskii [\onlinecite{PerPinsk}] we seek the distribution function $f_{\bm p}$ of electron quasimomentum ${\bm p}$ in the form
\[
f_{\bm p} = f_0 (\varepsilon_{\bm p}) + f_1 ({\bm p}) {\rm e}^{{\rm i} \omega t} + f_1^* ({\bm p}) {\rm e}^{-{\rm i} \omega t}\:,
\] 
expanding the functions $f_0$ and $f_1$ as follows
\begin{eqnarray} \label{f0f1}
&&f_0 (\varepsilon_{\bm p}) = \Phi(\varepsilon_{\bm p}) + A(\varepsilon_{\bm p}) + a_{\alpha} v_{\alpha} + C_{\alpha \beta} v_{\alpha \beta} + b_{\alpha \beta \gamma} v_{\alpha \beta \gamma}\:,
\nonumber\\ 
&&f_1({\bm p}) = A'(\varepsilon_{\bm p}) + R_{\alpha} v_{\alpha} + C'_{\alpha \beta} v_{\alpha \beta}\:, 
\end{eqnarray} 
where $\varepsilon_{\bm p} = p^2/m, {\bm v} = {\bm p}/m$ are the electron energy and velocity,
\begin{eqnarray}
&&v_{\alpha \beta} = v_{\alpha} v_{\beta} - \frac{1}{d} \delta_{\alpha \beta} v^2\:, \nonumber \\ &&v_{\alpha \beta \gamma} = v_{\alpha} v_{\beta} v_{\gamma}
- \frac{v^2}{d + 2} \left( v_{\alpha} \delta_{\beta \gamma} +  v_{\beta}  \delta_{\alpha \gamma} + v_{\gamma} \delta_{\alpha \beta}\right)\:, \nonumber
\end{eqnarray} 
and the dimension of space $d$ is 2 or 3. The other notations are 
\[
\Phi(\varepsilon) = \frac{1}{\exp{[(\varepsilon - \mu)/k_B T]} + 1}  
\]
\\
for the equilibrium energy distribution of electrons ($k_B$ is the Boltzmann constant), $A(\varepsilon)$ for the non-equilibrium correction to $\Phi$ that does not contain coordinate derivatives of the field and satisfies the equation
\begin{equation} \label{SA}
\frac{2 e}{m}{\rm Re} \left[ E^*_{\alpha} \left( R_{\alpha} + \frac{2\varepsilon}{d} \frac{\partial R_{\alpha}}{\partial \varepsilon}\right)\right] = S(A)
\end{equation}
under condition $\sum_{\bm p} A = 0$, $A'$ for the correction to the distribution function linear in the electric field, independent of the direction of ${\bm p}$ and satisfying the equation
\begin{equation} \label{Aprime}
- {\rm i} \omega A' + \frac{v^2}{d} \frac{\partial R_{\alpha}}{\partial x_{\alpha}} = S(A')\:,
\end{equation}
and $S(...)$ for the collision integral. Multiplying the left- and right-hand sides of (\ref{Aprime}) by the charge $e$ and summing over the quasimomentum and spin, we get the standard continuity equation
\[
- {\rm i} \omega e \delta N + \frac{\partial j_{1, \alpha}}{\partial x_{\alpha}} = 0\:,
\]
where ${\bm j}_1 = \sigma(\omega) {\bm E}$ and $\delta N = (2/V_d)\sum_{\bm p} A'$ are a current linear in the electric field and a local change of the electron density, respectively, ${\bm E}$ is the amplitude of high-frequency electromagnetic field, $V_d$ is the macroscopic volume in the $d$-dimensional space, and we introduce the conductivity at the frequency $\omega$:
$$\sigma(\omega) = \frac{e^2 N \tau}{m (1 - {\rm i} \omega \tau)}\:.$$

The generated electric current is related with the vector function ${\bm a}(\varepsilon)$ by
\begin{equation} \label{ja}
{\bm j} = \frac{2e}{d V_d} \sum\limits_{\bm p} v^2 {\bm a}(\varepsilon_{\bm p})\:.
\end{equation}
To the first order in gradients we find 
\begin{widetext}
\begin{eqnarray} \label{23}
&&\mbox{}\hspace{1 cm}a_{\alpha} = - \tau \left[ \frac{\partial \Phi}{\partial x_{\alpha}} + \frac{\partial A}{\partial x_{\alpha}} + \frac{2}{d+2} v^2 \left( 
\frac{\partial C_{\alpha \beta}}{\partial x_{\beta}} - \frac{1}{d} \frac{\partial C_{\beta \beta}}{\partial x_{\alpha}} \right) +
2 e {\rm Re} \left( \frac{\partial A' }{\partial \varepsilon} E^*_{\alpha} \right) \right.  \\ && \left. + \frac{4e}{m} \left(C'_{\alpha \beta}E^*_{\beta} - \frac{1}{d} C'_{\beta \beta}E^*_{\alpha} \right) + \frac{4}{d+2} e v^2 {\rm Re} \left(
\frac{\partial C'_{\alpha \beta}}{\partial \varepsilon} E^*_{\beta} - \frac{1}{d} \frac{\partial C'_{\beta \beta}}{\partial \varepsilon} E^*_{\alpha} \right) - \frac{2e}{cm} {\rm Re}[{\bm R} \tilde{\bm B}^* ]_{\alpha} \right] \:, \nonumber
\end{eqnarray}
\end{widetext}
where, for a three-dimensional system, $\tilde{\bm B}$ coincides with the magnetic field ${\bm B}$, while in a two-dimensional system $\tilde{B}_x = \tilde{B}_y = 0$ and $\tilde{B}_z = B_z$, the axis 
$z$ is perpendicular to the $xy$-plane which allows the free movement of electrons, 
\begin{eqnarray} 
{\bm R} &=& - \frac{e {\bm E} \tau}{1 - {\rm i} \omega \tau}\frac{\partial \Phi}{\partial \varepsilon}\:, \label{RCC''}\\
C_{\alpha \beta} &=& - e \tau {\rm Re} \left( E^*_{\alpha} \frac{\partial R_{\beta}}{\partial \varepsilon} + E^*_{\beta} \frac{\partial R_{\alpha}}{\partial \varepsilon}\right)\:, \label{RCC'}\\
C'_{\alpha \beta} &=& -  \frac{\tau}{2(1 - {\rm i} \omega \tau)} \left( \frac{\partial R_{\alpha} }{\partial x_{\beta}} + 
\frac{\partial R_{\beta} }{\partial x_{\alpha}}\right)\:.
\end{eqnarray}
When $d=3$, Eq.~(\ref{23}) goes into Eq.~(28) of Ref.~[\onlinecite{PerPinsk}]. The difference in factors of 2 or 4 in the terms in Eqs.~(\ref{23}) and (\ref{RCC'}), as compared with the similar terms in Ref.~[\onlinecite{PerPinsk}], occurs because of the difference (by a factor of 2) in the definition of the field amplitudes ${\bm E}, {\bm B}$ and correction $f_1$ in the present work and the article
by Perel and Pinskii.

For the two-dimensional electron gas one has
\begin{widetext}
\begin{eqnarray} \label{2D}
&&\mbox{}\hspace{0.5 cm}a^{(2D)}_{\alpha}= - \tau \left[ \frac{\partial \Phi}{\partial x_{\alpha}} + \frac{\partial A}{\partial x_{\alpha}} + \frac{1}{2} v^2 \left( 
\frac{\partial C_{\alpha \beta}}{\partial x_{\beta}} - \frac{1}{2} \frac{\partial C_{\beta \beta}}{\partial x_{\alpha}} \right) +
2 e {\rm Re} \left( \frac{\partial A' }{\partial \varepsilon} E^*_{\alpha} \right) \right. \\ && \left. + \frac{4e}{m} \left(C'_{\alpha \beta}E^*_{\beta} - \frac12 C'_{\beta \beta}E^*_{\alpha} \right) + e v^2{\rm Re} \left(
\frac{\partial C'_{\alpha \beta}}{\partial \varepsilon} E^*_{\beta} - \frac{1}{2} \frac{\partial C'_{\beta \beta}}{\partial \varepsilon} E^*_{\alpha} \right) - \frac{2e}{cm} [{\bm R} \tilde{\bm B}^* ]_{\alpha} \right] \:. \nonumber
\end{eqnarray}
\end{widetext}

Substituting the expression (\ref{23}) for ${\bm a}$ into the sum (\ref{ja}) we obtain, in the case where the rate of electron momentum scattering is independent of the energy and scattering angle, the following equation for the generated dc current density
\begin{widetext}
\begin{eqnarray} \label{PP2}
j_{\alpha} &=&  \frac{e \tau}{m} \left\{ - \frac{2}{d} \frac{\partial }{\partial x_{\alpha}} [N (\bar{\varepsilon} + \Delta \bar{\varepsilon})] - 2 \sigma'(\omega)\tau 
\frac{\partial }{\partial x_{\beta}}  \left( E_{\alpha} E^*_{\beta} +  E^*_{\alpha} E_{\beta} - \frac{2}{d} \delta_{\alpha \beta} \left|
{\bm E} \right|^2 \right) \right. \nonumber \\
&& \left. + \frac{2}{\omega} {\rm Im} \left[ E^*_{\alpha} \frac{\partial }{\partial x_{\beta}} \left( \sigma E_{\beta} \right) \right] + 
\frac{2}{c} {\rm Re} \biggl( \sigma [ {\bm E} \times \tilde{\bm B}^*]_{\alpha} \biggr) \right\}\:.
\end{eqnarray}\end{widetext}
Here $\sigma'(\omega) = {\rm Re}\{\sigma(\omega)\}$, $\bar{\varepsilon}(T)$ is the average electron kinetic energy in the equilibrium electron gas at the sample temperature $T$, $N \Delta \bar{\varepsilon}$ is a change of the electron energy proportional to the intensity of electromagnetic field. In the energy relaxation time approximation independent of the energy, in which case $S(A) = - A/\tau_{\varepsilon}$, one has
\[
N \Delta \bar{\varepsilon} = 2 \sigma'(\omega) \vert {\bm E}\vert^2\tau_{\varepsilon}\:.
\]
In the particular case of Boltzmann statistics, $\bar{\varepsilon} = (d/2) k_B T$ and $\Delta \bar{\varepsilon} = (d/2) k_B \Delta T$, where $\Delta T$ is the change of effective electron temperature caused by the radiation, and the first term in Eq.~(\ref{PP2}) takes the form
\[
- \frac{e \tau}{m} \frac{\partial }{\partial x_{\alpha}} [N k_B (T + \Delta T)] \:,
\]
in accordance with Refs.~[\onlinecite{PerPinsk}] and [\onlinecite{PRBlateral}].

For a plane electromagnetic wave propagating in a three-dimensional isotropic medium, the penultimate term in Eq.~(\ref{PP2}) vanishes, while the last term describes the high-frequency Hall effect, or the drag effect [\onlinecite{Barlow,Gurevich,PerPinsk1,Danish,Grin,Gibson}]. In the case of two-dimensional electron gas the indices $\alpha, \beta$  run through $x$ and $y$, the in-plane component ${\bm q}_{\parallel}$ of the light wave vector ${\bm q}$ is real, the first and second gradients in Eq.~(\ref{PP2}) disappear, the third term is different from zero and its sum with the fourth term gives the drag current
\begin{equation} \label{dragQW}
{\bm j} = \frac{2 e \tau}{m} \frac{{\bm q}_{\parallel}}{\omega}  \sigma'(\omega) \vert {\bm E}_{\parallel} \vert^2\:.
\end{equation}
With allowance for the dependence of momentum relaxation time $\tau$ on the electron energy, the photocurrent depends on the polarization state of the electric field component in the plane $(x, y)$, in particular, on the circular polarization [\onlinecite{DragQW}]. For charged particles with a linear dispersion law, for example, for electrons or holes in graphene, the polarization dependence does exists even neglecting the energy dependence of the relaxation time [\onlinecite{DragGraph}].

\section{Generation of current in FET}
Consider the generation of photovoltage by a plasmon wave propagating from the source ($x = 0$) to the drain ($x = L$) in the FET with a two-dimensional channel [\onlinecite{Shur},\onlinecite{Shur2}]. The two-dimensional electron gas is affected by the electric field
\[
{\bm E}(x,t) = - {\bm o}_x \frac{\partial U(x,t)}{\partial x} \:,
\]
where ${\bm o}_x$ is the unit vector in the $x$-axis direction, $U(x,t)$ is the electric potential determined by the boundary conditions
\[
U(0,t) = U_0 + U_a \cos{\omega t}\:,\: j(L,t) = 0\:,
\]
$U_0$ is the constant component of voltage between the gate and the channel related to the density $N$ by $N = CU_0$, with $C$ being the gate capacitance, the amplitude $U_a$ of the varying component depends on the excitation conditions. The alternating signal $U_1(x, t) = U(x, t) - U_0$ is a superposition of exponential functions
\begin{equation} \label{U1}
U_1(x,t) = {\rm e}^{- {\rm i} \omega t} \left( C_1 {\rm e}^{ {\rm i} kx} + C_2 {\rm e}^{-{\rm i} kx} \right) + \mbox{c.c.}\:,
\end{equation} 
where the wave vector is given by
\[ 
k = \frac{\omega}{s} \sqrt{1 + \frac{\rm i}{\omega \tau} }\:,\: s = \sqrt{\frac{e U_0}{m}}\:.
\]
\subsection{Semiinfinite channel}
In a long channel with $\exp{(-k'' L)} \ll 1$, the coefficient $C_2$ in Eq.~(\ref{U1}) is negligibly small, in this case $C_1 = U_a/2$ and
\begin{equation} \label{E1}
{\bm E}(x,t) = E_0  {\rm e}^{- {\rm i} \omega t + {\rm i} kx}{\bm o}_x + \mbox{c.c.}\:,\: E_0 = - {\rm i} k U_a/2\:.
\end{equation} 
Substituting this expression for the field into the equation (\ref{PP2}) for the photocurrent, we find for $d = 2$
\begin{equation} \label{current1}
j_x (x) = \frac{2e\tau}{m \omega} \sigma'(\omega) \vert E_0 \vert^2 \left[ k' +  k'' \omega (\tau + 2 \tau_{\varepsilon}) \right] {\rm e}^{- 2 k'' x}\:.
\end{equation}
The first term, proportional to the real part  $k'$ of the wave vector, is a current of the drag of electrons by the plasmon wave [\onlinecite{Popov, Popov2, Popov3}]. Indeed, in the process of  interaction of the wave with electrons the electron gas acquires the momentum per unit time,  per unit area
\[
W^{({\rm in})}_p = \frac{k'}{\omega} 2 \sigma'(\omega) \vert E_{0}\vert^2 {\rm e}^{- 2 k'' x}\:,
\]
which is being lost due to scattering: the ``inflow'' $W^{({\rm in})}_p$ is compensated by the ``outflow''
\[
W^{({\rm out})}_p = N \bar{p}_x/\tau \:,
\]
\\
where $\bar{p}_x$ is the mean value of the quasimomentum component $p_x$ in the electron gas. Finding a value of $\bar{p}_x$ from the balance equation $W^{({\rm in})}_p = W^{({\rm out})}_p$ and taking into account that the current $j_x$ equals $e N \bar{p}_x/m$, we obtain the first term in (\ref{current1}). The remaining term in this equation is proportional to the extinction coefficient $2 k''$. Moreover, the contribution proportional to $\tau_{\varepsilon}$ is related to the gradient of the electron-gas heating.

Under steady-state conditions, the external current $j_x(x)$ is compensated by the countercurrent $- \sigma(0) d U(x)/dx$ created by the static electric potential $U(x)$ of the shifted
charges. Taking into account the identities
\[
\vert k \vert^2 = \left( \frac{\omega}{s}\right)^2 \frac{\sqrt{1 + (\omega \tau)^2}}{\omega \tau}\:,\:
\frac{k'}{k''} = \sqrt{1 + (\omega \tau)^2} + \omega \tau\:,
\]
one obtains for the emf
\begin{equation} \label{photovoltage}
\Delta U = \frac{U_a^2}{4U_0} 
\left[ 1 + \frac{ 2\omega (\tau + \tau_{\varepsilon}) }{ \sqrt{1 + \omega^2 \tau^2}} \right]\:.
\end{equation}

In Ref.~[\onlinecite{Shur2}] the formula for the emf in FET is derived from the Euler equation written in the form 
\begin{equation} \label{I1}
\frac{\partial V}{\partial t} + V \frac{\partial V}{\partial x} + \frac{e}{m} \frac{\partial U(x,t)}{\partial x} + \frac{V}{\tau} = 0\:,
\end{equation}
where $V$ is the local electron velocity. The result in the form of Eqs.~(23) and (34) in Ref.~[\onlinecite{Shur2}] does not contain the energy relaxation time $\tau_{\varepsilon}$. The term in (\ref{photovoltage}) due to the heating is obtained from the Euler equation if a contribution from the pressure is included into this equation by adding into the left-hand side of Eq.~(\ref{I1}) the term $\rho^{-1}(\partial P/\partial x)$, in which the density $\rho$ equals $N m$ and the pressure $P$ is $(2/d) N (\bar{\varepsilon} + \Delta \bar{\varepsilon})$. In the case of nondegenerate statistics we obtain the equation $P = N k_B (T + \Delta T)$ consistent with the classical equation relating the equilibrium pressure, concentration and temperature [\onlinecite{Landau6},\onlinecite{Landau10}]. The summand
\[
\frac{1}{\rho} \frac{\partial P}{\partial x} \equiv \frac{2}{d} \frac{1}{N m} \frac{\partial}{\partial x}  [N (\bar{\varepsilon} + \Delta \bar{\varepsilon})]
\]
can be derived from the Boltzmann equation for the distribution function $f_{\bm p}$. Allowance for it in the Euler equation leads to an additional dc contribution
\[
- \frac{2}{d} \frac{\tau}{N m} \frac{\partial}{\partial x}  [N (\bar{\varepsilon} + \Delta \bar{\varepsilon})]
\]
to velocity $V$ and the contribution to the electric current $j_x = eNV$ coinciding with the first term in Eq.~(\ref{PP2}).

It is worth to note that the current (\ref{current1}) in FET proportional to $k'$ and the current (\ref{dragQW}) in the quantum well are both currents of the drag of electrons by the field. However, unlike the bulk medium, the current (\ref{current1}) cannot be interpreted as a high-frequency Hall effect, since the plasmon wave propagating along the transistor channel is longitudinal and lacks a magnetic field.

The contribution to the current (\ref{current1}) proportional to $k'' \omega \tau$ comes from the second and third terms in Eq.~(\ref{PP2}); while using the Euler equation (\ref{I1}), it arises from the nonlinear term $V dV/dx = (dV^2/dx)/2$ and can be interpreted as a consequence of nonlinear electron convection [\onlinecite{Popov3}]. Under transient conditions, the times of establishment and relaxation of this contribution, as well as of the drag current, are determined by the momentum relaxation time $\tau$, whereas the heating generation mechanism is controlled by the energy relaxation time $\tau_{\varepsilon}$. This may allow one to separate the heating and non-heating mechanisms when detecting a radiation whose intensity is modulated at a frequency $\Omega$ small compared with the carrier frequency $\omega$ but comparable to the reciprocal time $\tau^{-1}_{\varepsilon}$.
\subsection{The channel of finite length}
Using the solution for the electric potential and the local electron velocity in the channel of a finite length $L$, see Ref.~[\onlinecite{Shur2}], and adding the contribution due to the electron heating, we obtain for the photo-emf
\begin{eqnarray} \label{photovoltage2}
\frac{\Delta U}{U_0} &=& \frac14 \left(\frac{U_a}{U_0}\right)^2 f(\omega)\:,\\ f(\omega) &=& 1 + \tilde{\beta} - \frac{1 +  \tilde{\beta} \cos{2k'L}}{\sinh^2{k'' L} + \cos^2{k' L}} \:, \nonumber
\end{eqnarray} 
where
\[
\tilde{\beta} = \left( 1 + \frac{\tau_{\varepsilon}}{\tau} \right)\frac{2 \omega \tau}{\sqrt{1 + (\omega \tau)^2}}\:.
\] \newpage
In a thin channel satisfying the condition $s \tau \gg L$, the frequency dependence $f(\omega)$ has a resonance character at the frequencies $n \omega_0$, where $\omega_0 = \pi s/(2L)$ and $n$ are odd numbers 1, 3, 5$\dots$ In the vicinity of this frequency one has
\[
f(\omega) \approx \left( 1 + 2 \frac{ \tau_{\varepsilon} }{\tau} \right) \frac{(2 s \tau/L)^2}{4 \tau^2 (\omega - n \omega_0)^2 + 1} \:.
\]

\section{Conclusion}
We have derived an expression for the dc electric current generated by a high-frequency electromagnetic field in a two-dimensional electron gas. Both the drag effect of electrons by the field and
the appearance of a current due to the inhomogeneous heating of the carriers by the field are taken into account. The contribution of the second effect is limited by the energy relaxation time of the electron gas and can significantly exceed the current calculated neglecting the heating. Under the action of a modulated radiation on the two-dimensional electron gas, e.g. under conditions of the experiment [\onlinecite{Kukushkin}], a possibility arises to separate the heating and non-heating mechanisms of the electric-current generation.\\

\paragraph*{\textbf{Acknowledgement.}} The author is grateful to L.E.~Golub and V.V.~Popov for helpful discussions of the manuscript. Financial support of the Russian Science Foundation Grant 14-12-01067 and President Grant NSh-1085.2014.2 is gratefully acknowledged.

\end{document}